\documentclass[preprint,12pt,nonatbib]{elsarticle}

\usepackage{amssymb}
\usepackage{amsmath}
\usepackage[utf8]{inputenc}
\DeclareUnicodeCharacter{202F}{ }
\usepackage{tabularx}
\usepackage{lipsum}
\usepackage{array}
\usepackage{multirow}
\usepackage{booktabs} % for nicer tables
\usepackage{siunitx}  % for scientific notation & units
\usepackage{comment}
\usepackage{placeins}
\usepackage{arydshln} 
\usepackage{hyperref}
\usepackage[backend=biber, sorting=none, url=false, doi=true]{biblatex}
\newcolumntype{Y}{>{\centering\arraybackslash}X}

\journal{Acta Materialia}

\begin{document}

\begin{frontmatter}

%% Title, authors and addresses

%% use the tnoteref command within \title for footnotes;
%% use the tnotetext command for theassociated footnote;
%% use the fnref command within \author or \affiliation for footnotes;
%% use the fntext command for theassociated footnote;
%% use the corref command within \author for corresponding author footnotes;
%% use the cortext command for theassociated footnote;
%% use the ead command for the email address,
%% and the form \ead[url] for the home page:
%% \title{Title\tnoteref{label1}}
%% \tnotetext[label1]{}
%% \author{Name\corref{cor1}\fnref{label2}}
%% \ead{email address}
%% \ead[url]{home page}
%% \fntext[label2]{}
%% \cortext[cor1]{}
%% \affiliation{organization={},
%%             addressline={},
%%             city={},
%%             postcode={},
%%             state={},
%%             country={}}

\title{Rapid Primary Radiation Damage Resistance Assessment of Precipitation-Hardened Cu Alloys} %% Article title

\affiliation[label1]{organization={Plasma Science and Fusion Center (PSFC)},
%%             addressline={},
             city={Cambridge},
             postcode={02139},
             state={MA},
             country={USA}}
\affiliation[label2]{organization={Department of Materials Science \& Engineering},
%%             addressline={},
             city={Cambridge},
             postcode={02139},
             state={MA},
             country={USA}}
\affiliation[label3]{organization={Department of Nuclear Science \& Engineering},
%%             addressline={},
             city={Cambridge},
             postcode={02139},
             state={MA},
             country={USA}}
\author[label1,label2]{Elena Botica-Artalejo}
\author[label1]{Gregory Wallace}
\author[label1,label3]{Michael Short}

%%
%% Abstract
\begin{abstract}
%% Text of abstract
This study establishes a direct correlation between in situ irradiation-induced property changes measured by transient grating spectroscopy (TGS) and the resulting microstructural damage in Cu–Cr–Ta alloys. Thin films fabricated by physical vapor deposition were irradiated with 6.6 MeV $Cu^{+3}$ ions up to 25 DPA, while TGS continuously monitored the evolution of surface acoustic wave (SAW) frequency and thermal diffusivity. Post-irradiation transmission electron microscopy (TEM) was used to quantify void formation as a metric of accumulated radiation damage. A pronounced decrease in SAW frequency was observed some seconds after the onset of irradiation, and it was found to correlate strongly with the final void density. Vacancy MEB calculations propose that the small decrease in SAW frequency is associated with the low population of mobile vacancies, promoting defect recombination and decreasing void formation. This relationship enables early prediction of relative radiation damage resistance within minutes of irradiation, substantially reducing the time required compared to conventional irradiation and post-characterization routes, allowing rapid screening of multiple compositions. We were able to test this method with 3 compositions of the Cu-Cr-Ta system. More generally, this in situ approach provides an efficient pathway for accelerating the development of materials for radiation environments.
\end{abstract}

%%Graphical abstract
\begin{graphicalabstract}
\includegraphics[width=\textwidth]{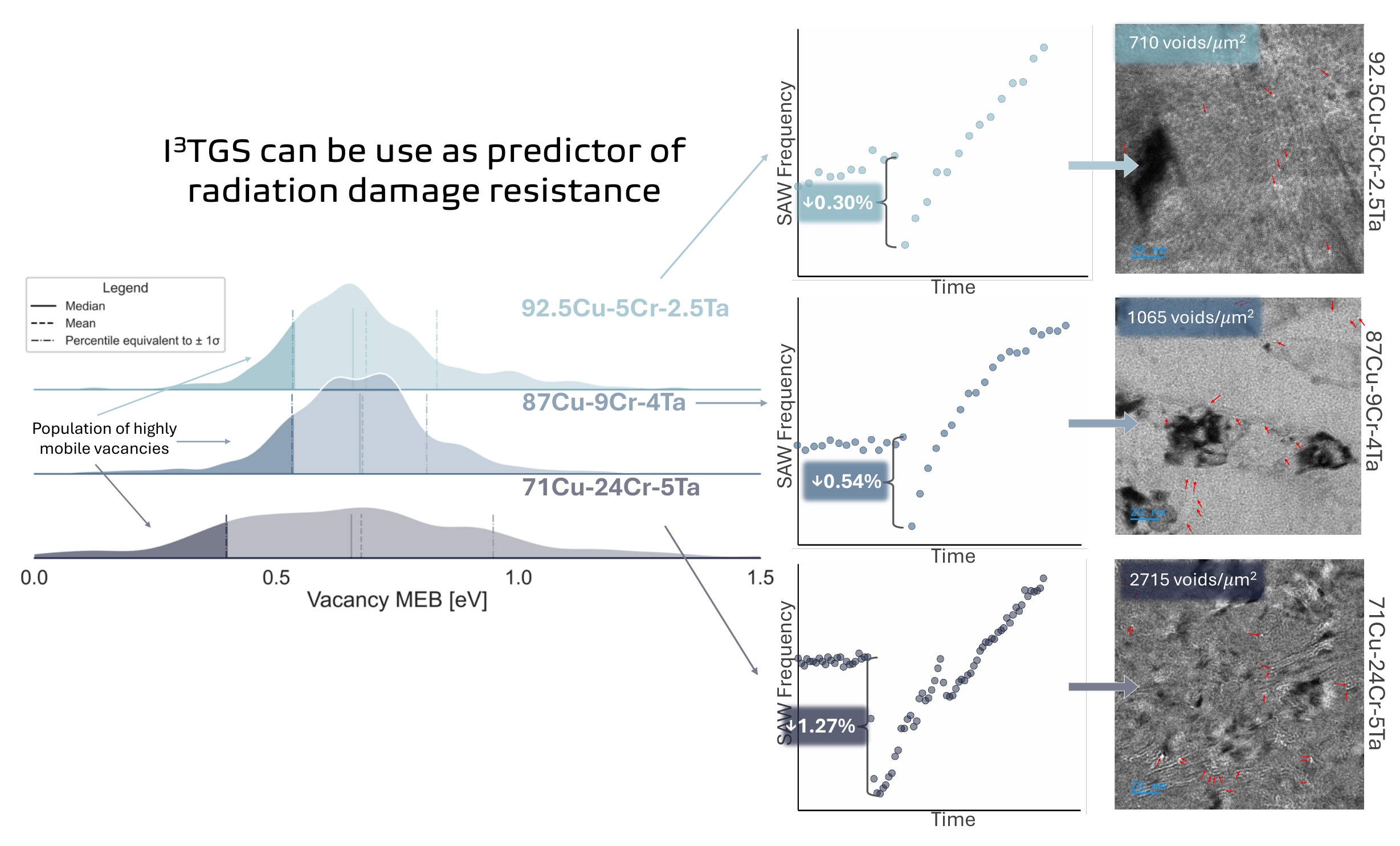}
\end{graphicalabstract}

%%Research highlights
\begin{highlights}
\item In situ transient grating spectroscopy reveals an immediate SAW frequency response to irradiation that is able to provide relative comparison of final void accumulation between different samples.
\item Vacancy MEB distributions explain the susceptibility to form voids, by the population of mobile vacancies.
\item This methodology enable rapid screening of radiation damage resistance materials, reducing irradiation and characterization times.
\end{highlights}

%% Keywords
\begin{keyword}
%% keywords here, in the form: keyword \sep keyword
\emph{In situ} Irradiation \sep TGS \sep Nuclear materials \sep Cu-alloys \sep GRCop-84
%% PACS codes here, in the form: \PACS code \sep code

%% MSC codes here, in the form: \MSC code \sep code
%% or \MSC[2008] code \sep code (2000 is the default)

\end{keyword}

\end{frontmatter}

%% Add \usepackage{lineno} before \begin{document} and uncomment 
%% following line to enable line numbers
%% \linenumbers

%% main text
%%

%% Use \section commands to start a section
\section{Introduction}
Fusion technology presents exceptional challenges across multiple scientific disciplines, particularly in materials science, where it imposes extreme demands on structural and functional components. All plasma-facing components (PFCs) are subject to intense fluxes of neutrons, ions, and thermal energy. To perform reliably in this environment, PFCs must meet a stringent set of common requirements, including low activation, minimal sputtering erosion, low tritium retention, a high melting point, and stable thermo‑mechanical properties. Beyond these baseline criteria, specialized applications introduce further constraints. A prominent example is the radio‑frequency (RF) antenna, used to heat plasma to ignition temperatures in tokamaks, which must satisfy additional performance specifications. These RF antennas must exhibit high electrical conductivity to mitigate resistive losses and demonstrate low susceptibility to arcing under high‑power conditions \cite{wallace_multiphysics_2021}. As these components are designed for continuous, maintenance‑free operation over extended durations—spanning months or years—their ability to withstand the fusion environment without degradation is paramount.

The development of viable lower hybrid current drive (LHCD) launchers is critically constrained by the lack of a material that satisfies all operational, thermal-mechanical, and radiological requirements simultaneously \cite{wallace_ultra-rapid_2022}. A leading candidate is the precipitation-hardened copper alloy GRCop-84 (Cu–8Cr–4Nb, at.\%), which excels in several key performance areas, and has been used for construction of the LHCD~\cite{wukitch_diii-d_2023} and reflectometer~\cite{leppink_high-field_2025} antennas in the DIII-D tokamak. Its strength derives from a dispersion of fine, thermally stable Cr$_2$Nb (C15 Laves phase) precipitates within the $\mu m-nm$ size range, which resist coarsening up to $800^\circ C$ and pin Cu grain boundaries, maintaining microstructural integrity \cite{ellis_grcop-84_2005, minneci_microstructural_2021}. Consequently, GRCop-84 exhibits high strength, low creep, and excellent thermal and electrical properties—with thermal diffusivity and electrical resistivity ($\sim 8.3×10^{-5} m^2/s$ and $\sim 3 \mu \Omega\cdot cm$ at $100^\circ C$, respectively) comparable to oxygen-free copper \cite{ellis_thermophysical_2000, yang_thermal_2014, siu_thermal_1976}. These attributes make it an exceptional candidate for RF components \cite{seltzman_surface_2020, seltzman_rf_2020}.

However, its viability for fusion is ultimately limited by the 4 at.\% niobium content, which presents a severe radiological drawback. For a material to qualify as low-level waste (LLW), a prerequisite for the social license of fusion power \cite{hoedl_achieving_2022}, it must achieve a waste disposal rating (WDR) below 1. The WDR is the sum of the ratios between an element’s concentration divided by that element’s maximum allowable concentration after irradiation, considering its potential to become dangerously radioactive. The Nb in GRCop-84 leads to a calculated WDR of $\sim 2400$ after 1.7 years of operation \cite{klueh_impurity_2000}, orders of magnitude above the acceptable threshold and far exceeding the established limit of 0.4–9 ppm for Nb in first-wall materials \cite{fetter_long-term_1990}.

This fundamental discrepancy creates a clear imperative: to develop Nb-free copper alloys that retain the beneficial C15 Laves phase precipitate strengthening while achieving a low WDR. The key to this new material system lies in replicating or replacing the stabilizing Cr$_2$Nb intermetallics. Fabrication of such microstructures typically requires rapid solidification techniques,  such as laser power bed fusion (LPBF) of gas atomized GRCop-84 powder, or even vapor deposition techniques such as physical vapor deposition (PVD) or chemical vapor deposition (CVD), to achieve the necessary fine-scale precipitates while avoiding detrimental grain growth or uneven distribution of the precipitates. 

Recent efforts to develop alloys with reduced niobium content have achieved some success. For instance, Perrin et al. \cite{perrin_microstructure_2024} investigated the Cu-Cr-Nb-Zr system to enhance the strength and creep life of ITER's baseline Cu-Cr-Zr alloy, demonstrating that Laves phase precipitates improve mechanical properties without degrading thermal and electrical conductivity post-irradiation. Nevertheless, the Nb content remains a limiting factor. Other candidate systems, such as Cu-Ni-Be \cite{zinkle_evaluation_2014} and Cu-Al$_2$O$_3$ \cite{simos_200_2019}, are constrained by their maximum operating temperatures. Among elements capable of forming C15 Laves phases with chromium, tantalum (Ta) emerges as a particularly promising alternative \cite{kumar_structural_2000}. Supporting this, Li et al. \cite{li_achieving_2025} recently demonstrated that a Cu-rich Cu-Cr-Ta alloy offers high-temperature mechanical property enhancements respect  Cu-Cr, taking advantage of the Laves phase precipitates, establishing this system as a compelling candidate for further investigation.

The present study examines the Cu-Cr-Ta alloy system as a prospective Nb-free material for RF antenna applications in fusion environments. To evaluate primary radiation damage resistance, a series of compositions were subjected to in situ heavy-ion irradiation. During irradiation, the materials' elastic properties were monitored in real time via transient grating spectroscopy (TGS), tracking the values of the surface acoustic wave frequency. This approach provides a high-throughput screening method, using the correlation of the immediate SAW frequency shift observed at beam onset with the material's propensity for vacancy creation, offering a rapid indicator of radiation tolerance. Findings from this in situ analysis were directly corroborated by post-irradiation examination using transmission electron microscopy (TEM), and the vacancy migration energy barrier (MEB) distributions from each system supported the findings.

\section{Methods}
\subsection{High-throughput material generation}
\begin{figure}
    \centering
    \includegraphics[width=\textwidth]{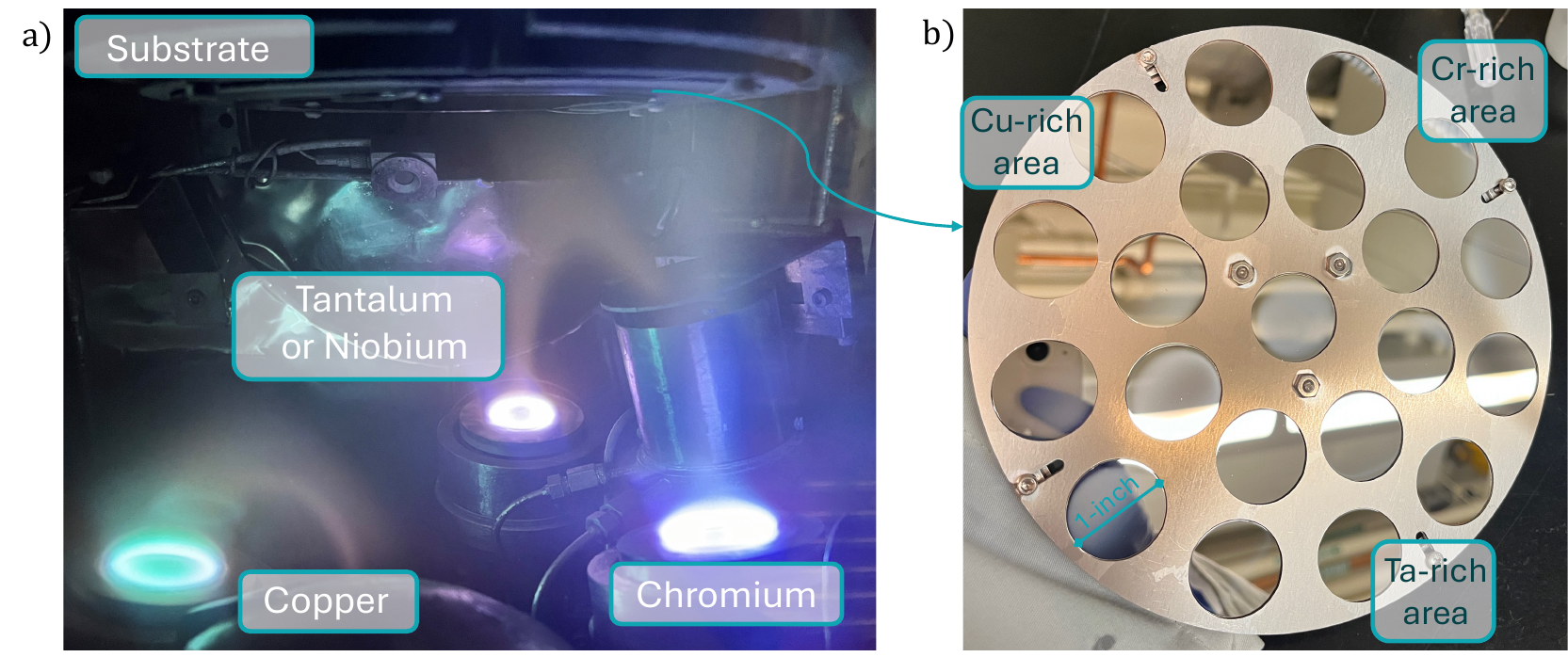}
    \caption{\small a)Combinatorial deposition of Cu-Cr-Ta alloys. b)Substrate holder with Si wafers after deposition.}
    \label{fig:PVD_lay}
\end{figure}

The material of study from this work has been thick films ($\sim 4-5 \mu m$) of combinatorial physically vapor deposited Cu alloys, changing its composition in a planar configuration along several 1-inch Si wafers, as depicted in Figure \ref{fig:PVD_lay}. Cu-Cr-Ta combinatorial thick films were deposited using physical vapor deposition in a Kurt J. Lesker PVD 75 PRO Line, enabling 2D gradient compositional films by independently controlling three sputtering targets favoring Cu-rich films (deposition parameters can be seen in Table \ref{tab:deposition_parameters}). To ensure a homogeneous deposition conditions between different elements, the three sputtering targets were arranged symmetrically at 120° separations, with each gun fixed at an identical inclination toward the substrate to create different concentrations patterns. The power for each target was adjusted to vary the amount relative concentration of each element. The following high‑purity Kurt J. Lesker targets were employed: Cu (99.99\%), Cr (99.95\%) and Ta (99.95\%). Impurity certifications are available in the associated data repository. Prior to deposition, a Ti (99.995\% pure) adhesion layer was applied to enhance substrate bonding and act as an oxygen getter, reducing contamination. The films were grown on 1-inch p-type (B-doped) semi-prime quality silicon wafers with $R_a < 0.5\mu m$. A total of 21 wafers are obtained per deposition, with a variation between 0.5-2 at.\% depending on the element considered.

\begin{table}[h]
\centering
\caption{Deposition parameters for Cu-Cr-Ta thick films.}
\label{tab:deposition_parameters}
\small
\begin{tabularx}{\textwidth}{>{\raggedright\arraybackslash}X c}
\toprule
\textbf{Parameter} & \textbf{Value}  \\ 
\midrule
Vacuum level prior to deposition (Torr) & \(7.4 \times 10^{-6}\)\\ 
Chamber pressure at deposition (Torr)   & \(6.2 \times 10^{-4}\) \\ 
Sputtering time (min)                   &150\\ 
Gas flow (sccm)                         & 20\\ 
Gas species                             & {Argon (99.999\%)} \\ 
Temperature (\(^\circ\)C)               & {20--25} \\ 
Cu target power \& size (Watts/inch/type)    & 150/2/DC \\ 
Cr target power \& size (Watts/inch/type)    & 80/2/RF  \\ 
Ta target power \& size (Watts/inch/type)     & 30/2/DC  \\
Ti target power \& size (Watts/inch/type)    & {60/2/DC} \\ 
Ti sputtering time (min)                & {60} \\ 
\bottomrule
\end{tabularx}
\end{table}

\subsection{Microscopy characterization}
The elemental composition on each wafer was obtained by energy-dispersive X-ray spectroscopy (EDS) using a Zeiss Gemini 450 scanning electron microscope (SEM). The gun energy was set to 25kV, to excite the characteristic X-ray lines and sampling a reasonable depth of the material. The working distance was maintained at 8.5 nm for optimal X-ray detection efficiency. In EDS analysis, carbon and oxygen spectral peaks were deconvolved to determine relative elemental concentrations while accounting for potential surface contaminants. A summary of sample compositions and the tests performed on each is provided in Table \ref{tab:chem_comp}.

\begin{table}[h]
    \centering
    \caption{Chemical composition and corresponding batch for Cu-Cr-Ta thick films.}
    \label{tab:chem_comp}
    \small
    \setlength{\tabcolsep}{3pt}
    \begin{tabular}{c c c c }
    \toprule
    \textbf{SampleID} & \textbf{Cu (at.\%)} & \textbf{Cr (at.\%)} & \textbf{Ta (at.\%)}\\
    \midrule
    044\_04 & 92.6 & 4.9 & 2.5\\
    044\_15 & 86.9 & 9.2 & 3.8\\
    044\_18 & 70.4 & 24.3 & 5.3\\ 
    \bottomrule
    \end{tabular}
\end{table}

Transmission electron microscopy (TEM) was utilized to characterize the damage induced by the ion beam. Void structures were imaged in conventional TEM mode using a JEOL F200 microscope operating at an accelerating voltage of 200 kV. A $30\mu m$ objective aperture was employed to mitigate contrast from other microstructural features. Subsequent image analysis, performed with Fiji software, enabled the quantification of void density through manual counting and sizing, using the ROI Manager function under the Process menu of the software. 

Complementary chemical analysis was conducted via energy-dispersive X-ray spectroscopy (EDS) on both irradiated and unirradiated regions to identify radiation-induced compositional and microstructural changes. Finally, electron energy loss spectroscopy (EELS) was performed in some of the TEM lamellas to know the thickness of them. The EDS and EELS analysis were done using the Thermo Fisher Scientific (TFS) Themis Z G3 aberration-corrected scanning transmission electron microscope (STEM) with a voltage of 300kV. To prepare the lamellas for TEM analysis the FEI Helios 660 focus ion beam (FIB) was used. Samples were progressively milled down up to $40-50\mu m$ by reducing the ion current from 30kV to 2kV. 

Prior to TEM inspection a surface cleaning was performed using a Fischione 1051 TEM mill to eliminate any Ga damage and surface oxides, which readily form on copper.

\subsection{In situ ion irradiation coupled with transient grating spectroscopy}
\begin{figure}
    \centering
    \includegraphics[width=0.9\textwidth]{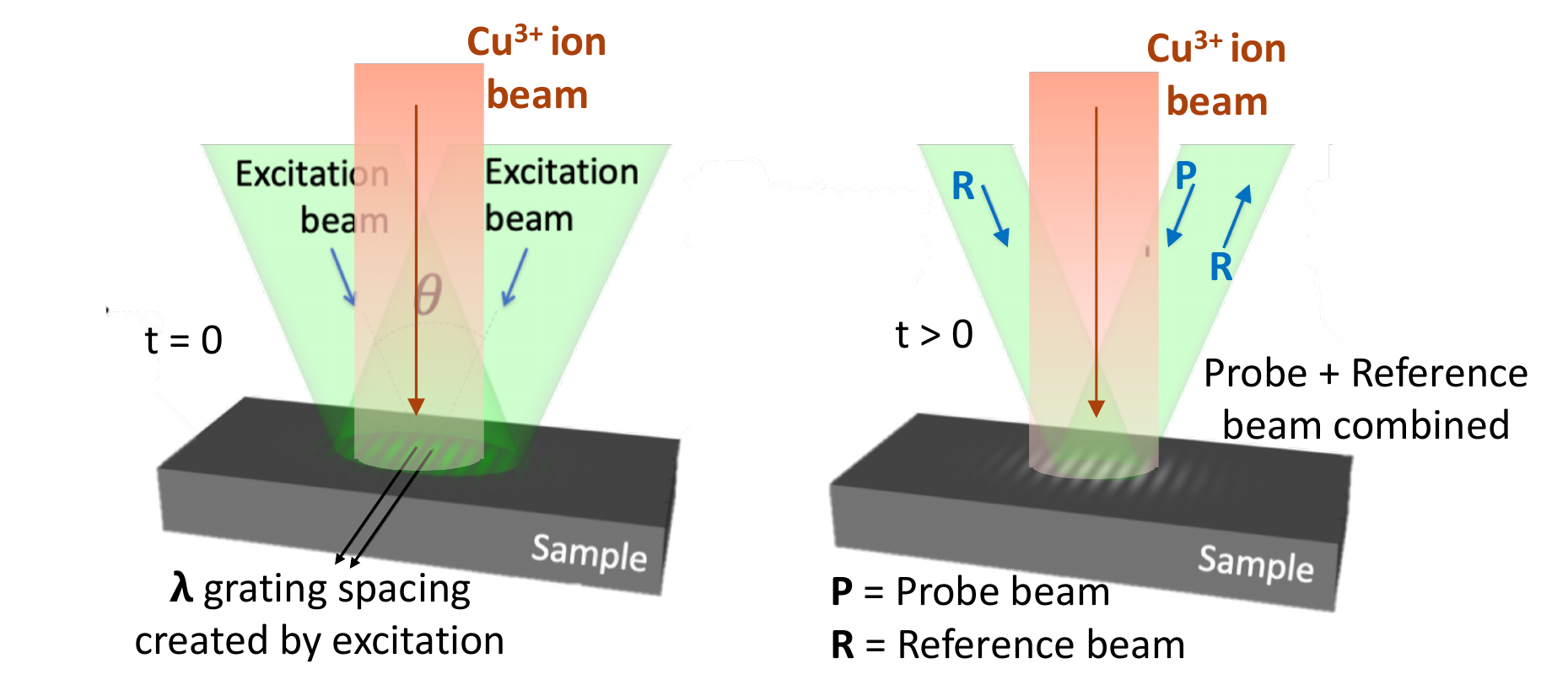}
    \caption{\small Schematic of TGS coupled with in situ ion irradiation. Adapted from \cite{hofmann_transient_2019}.}
    \label{fig:TGS_phy}
\end{figure}

Transient grating spectroscopy (TGS) is a non-contact, laser-based method used to quantitatively measure a material's elastic properties and thermal diffusivity at controlled depths \cite{hofmann_transient_2019}. The technique provides localized measurements of the thermo-elastic response with a spatial resolution determined by the focal width of the probe laser. Figure \ref{fig:TGS_phy} illustrates the conceptual setup when TGS is coupled with ion beam irradiation.
A pump laser ($1kHz$, $532nm$, $\sim 500ps$ pulses) excites the material surface, generating a grating pattern with an on-sample power of $7.4mW$. The grating spacing determines the maximum depth for probing thermo-elastic properties. A continuous-wave probe laser (577 nm) then monitors the surface excitation via first-order diffraction, with an intensity of approximately $127kW m^{-2}$. Table \ref{tab:tgs_param} details all relevant TGS parameters.
The probe beam measures the surface acoustic wave (SAW) frequency, which relates to elastic properties, and the signal decay provides the thermal diffusivity. The total signal intensity, ($I_{tot}(t)$), is described by:

\begin{equation}
I_{tot} (t)= A \Big[ 
\mathrm{erfc} \Big( q_0 \sqrt{\alpha_x t} + \frac{\beta}{\sqrt{t}} e^{-q_0^2 \alpha_x t} \Big)
\Big]
+ B \Big[ \sin(2 \pi f_{SAW} t + \Theta) e^{-t/\tau} \Big] + C
\end{equation}

where \(q_0 = 2\pi/\Lambda\) is the grating wavenumber originated from the $\Lambda [m]$ TGS grating space (in the present system configuration the focusing length is 1:1 through the phase mask, making $\Lambda$ being indistinguishable from the wavelength $\lambda$), $\alpha_x [m^2s^{-1}]$ is the in-plane thermal diffusivity, $\beta [s^{1/2}]$ represents the ratio of displacement-to-reflectivity contributions, $f_{SAW}[Hz]$ is the SAW frequency, $\Theta$ is the phase, and $\tau [s]$ is the decay time. Constants \(A\), \(B\), and \(C\) are determined through nonlinear fitting of experimental data \cite{wylie_accelerating_2025}.

The SAW frequency is related to elastic properties via the Rayleigh wave equation \cite{royer_rayleigh_1984}:

\begin{equation}
c_{r} = f_{SAW} \lambda = \sqrt{\frac{E}{\rho}} \, 
\end{equation}

where $\lambda [m]$ is the spatial period of the surface grating,  $E [Pa]$ is Young's modulus, and $\rho [kg/m^3]$ is the material density. For isotropic materials, this provides a quantitative measure, allowing the SAW frequency to be expressed as:

\begin{equation}
f_{SAW} = \frac{1}{\lambda} \sqrt{\frac{E}{\rho}}
\label{eq:SAW_to_E}
\end{equation}

For anisotropic materials, such as most polycrystals, absolute values are more appropriate, given the technique's spot size relative to grain dimensions. In this case, as we deal with a nanocrystalline microstructure, it can be assumed that we are obtaining an average value from different grains, as the spot size area is 10$^4$ times greater than the average grain size of the samples presented on this work.

A 0.2\% neutral density filter was employed to reduce the on-sample pump power to $5.89 mW$. A di-homodyne phase collection geometry was used \cite{dennett_thermal_2018}, employing two phase-locked probe beams ($\phi = \pm \tfrac{\pi}{2}$) to stabilize the signal against phase fluctuations. The current experimental setup is detailed in \cite{wylie_accelerating_2025}. The probing depth can be adjusted by changing the grating spacing, $\Lambda$, with nominal sensitivities of $\lambda/\pi$ for thermal properties \cite{kading_transient_1995} and $\lambda/2$ for SAW signals \cite{royer_rayleigh_1984}. A spacing of $\lambda=3.4 \mu m$ was selected to match the ion-damaged region depth. Daily calibrations using a single-crystal tungsten {100} sample ensured accurate grating spacing values, accounting for minor variations ($< \pm 0.1 \mu m$) due to ambient conditions.

Time resolution was optimized and settled on 2000 traces per 10 seconds, which improved temporal resolution without compromising data quality. A custom MATLAB code processed raw TGS data to extract SAW frequency and thermal diffusivity \cite{short_lab_github_2025}. Both raw and processed data are available in this article’s data repository.

\begin{table}[h]
    \centering
    \begin{tabular}{c c}
    \toprule
        \textbf{Parameter} & \textbf{Value} \\
        \hline  
        Pump wavelength & 532 nm \\
        Pump energy per pulse & 10 $\mu$J \\
        Pump pulse duration & $<$500 ps \\
        Pump spot size (on-sample diameter) & 150 $\mu$m \\
        Pump firing frequency & 1 kHz \\
        Pump on-sample power & 7.36 mW \\
        Probe wavelength & 671 nm \\
        Probe frequency & Continuous Wave \\
        Probe spot size (on-sample diameter)& 90 $\mu$m \\
        Probe on-sample power & 4.6 mW \\
        Detector bandwidth & 50 kHz - 1 GHz (3 dB) \\
        Oscilloscope bandwidth & 5 GHz \\
        Averaged traces per second & 2000 \\
        Calibrated on-sample grating spacing & 3.4 $\mu$m \\
        Sample roughness& $R_a = 10$ \AA \\
        \hline
    \end{tabular}
    \caption{Detailed summary of TGS system parameters. Spot sizes reported as $2\sigma$ Gaussian widths.}
    \label{tab:tgs_param}
\end{table}

The TGS system was mounted pointing to a chamber viewport, enabling in situ monitoring during irradiation. Samples were rotated 45$^\circ$ relative to the ion beam to maintain perpendicular alignment with the TGS system, required for di-homodyne signal formation.

Irradiations were performed at the Cambridge Laboratory for Accelerator Surface Science (CLASS) at the MIT Plasma Science and Fusion Center, using tandem ion accelerator at 1.65 MV. Experiments employed Cu$^{3+}$ ions at 6.6 MeV at around 250$^\circ C$. Flux rates ranged from $5\times 10^{16} ions/m^2\cdot s$ to $6\times 10^{16} ions/m^2\cdot s$. Beam current was monitored with a picoammeter connected to the sample holder, isolated electrically from the rest of the beamline. A 2 $mm$ aperture focused the beam, producing an ellipsoidal spot of 4.44 $mm^2$ on the angled sample. Prior to experiments, beam alignment with the TGS spot was verified using a silver-activated zinc sulfide coating on a ceramic substrate of 1 $mm$ thick laid on top of the sample.

Ion penetration depth and damage profiles were simulated using SRIM \cite{ziegler_srim_2010}, following the procedure indicated in \cite{stoller_use_2013} with modified displacement energies of 30 eV for Cu, 40 eV for Cr and 90 eV for Ta, according to \cite{astm_astm_2023}, obtaining a penetration depth of $2\mu m$.

\subsection{CALPHAD calculations}
CALPHAD calculations were performed using the software Thermo-Calc 2024b with the TCHEA7 database for high entropy alloys. Thermal conductivity and electrical conductivity calculations were performed, ranging the composition from pure Cu to pure Cr$_2$X, replacing X with different elements that form C15 laves phases as Nb does \cite{kumar_structural_2000}, like Zr, Ta, and Hf.

\subsection{Nudge Elastic Band calculations}
Climbing-image nudged elastic band (CI-NEB) calculations were performed with VASP to determine the vacancy migration energy barrier (MEB) in three Cu-based alloy systems. A 2x2x2 FCC supercell containing 64 atoms was employed. The supercells were generated using a Python script that randomly placed the constituent atoms according to the specified concentrations. Each system was initially relaxed to obtain a stable reference structure. Subsequently, another script was used to remove a Cu atom at the center of the supercell, creating a vacancy. The resulting configuration was relaxed without allowing changes to the supercell volume. Additional relaxed structures were generated by moving the vacancy position to each of the 12 nearest-neighbor (NN) positions relative to the original site. For each system, 20 random systems were generated.

The CI-NEB calculations were then performed using the relaxed microstructure with the central vacancy as the initial state and the vacancy-relaxed NN configurations as the final states of the migration pathway. A total of five intermediate images were used to construct the MEB profile. To account for the MEB value of a system, the calculated energy to move forward and backward was taken into account, implying 480 values for taking the average of the MEB for each of the alloy systems. 

Python scripts to create the supercell and NEB images, as well as, INCAR files for each calculation can be found in the Github repository of this manuscript.

\section{Results}
\subsection{Thermal conductivity calculations for Cu-based alloy systems}
A computational screening of candidate Cu–Cr–X systems was conducted using the CALPHAD method to assess thermal and electrical conductivity. The ternary element X was restricted to Zr, Ta, Nb, and Hf, all of which are known to form stabilizing C15 Laves phases with Cr \cite{kumar_structural_2000}, to obtain a precipitation-hardened alloy. To ensure a consistent comparison, the atomic composition was held equivalent to that of GRCop-84, with Nb replaced by each candidate.

\begin{figure}
    \centering
    \includegraphics[width=0.9\textwidth]{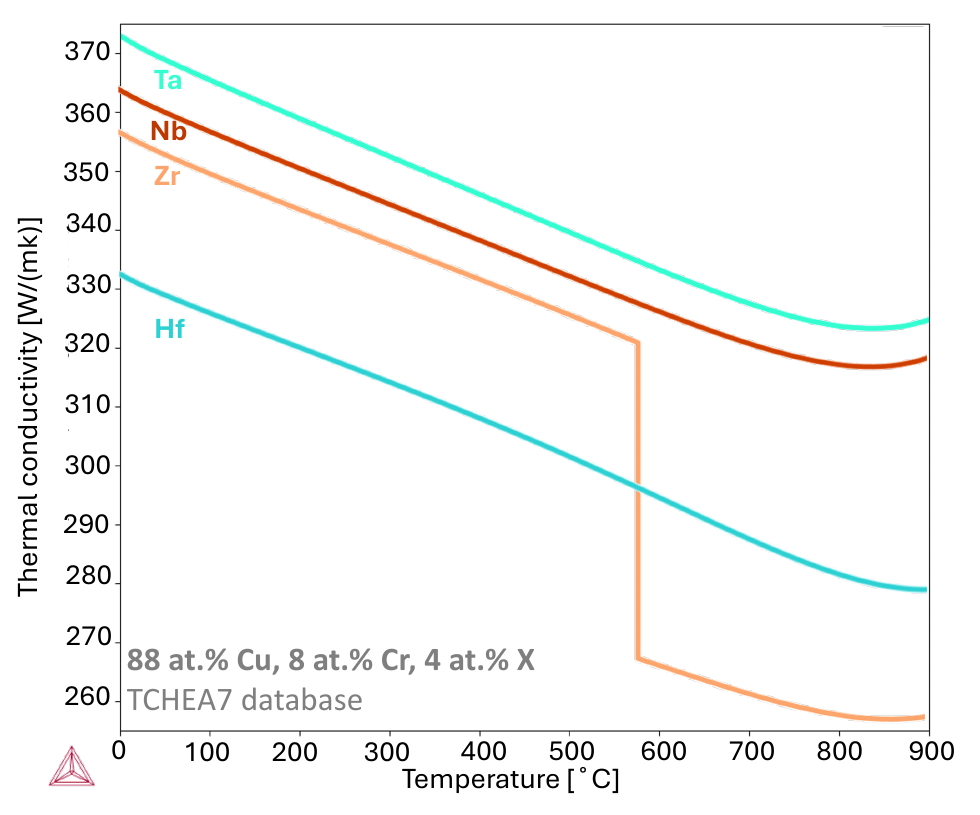}
    \caption{\small CALPHAD Thermal conductivity evaluation between different candidate elements in the 88 at.\%Cu-8 at.\%Cr-4at.\%X system. TCHEA7 database was employed.}
    \label{fig:TC}
\end{figure}

The results, presented in Figure \ref{fig:TC}, indicate that the Ta-substituted alloy possesses the highest calculated thermal conductivity. Through the Wiedemann–Franz relation \cite{yadav_analytic_2019}, this also implies superior electrical conductivity, projecting performance that may surpass the GRCop-84 benchmark.

\subsection{TGS monitoring during self-ion irradiation}
\begin{figure}
    \centering
    \includegraphics[width=0.9\textwidth]{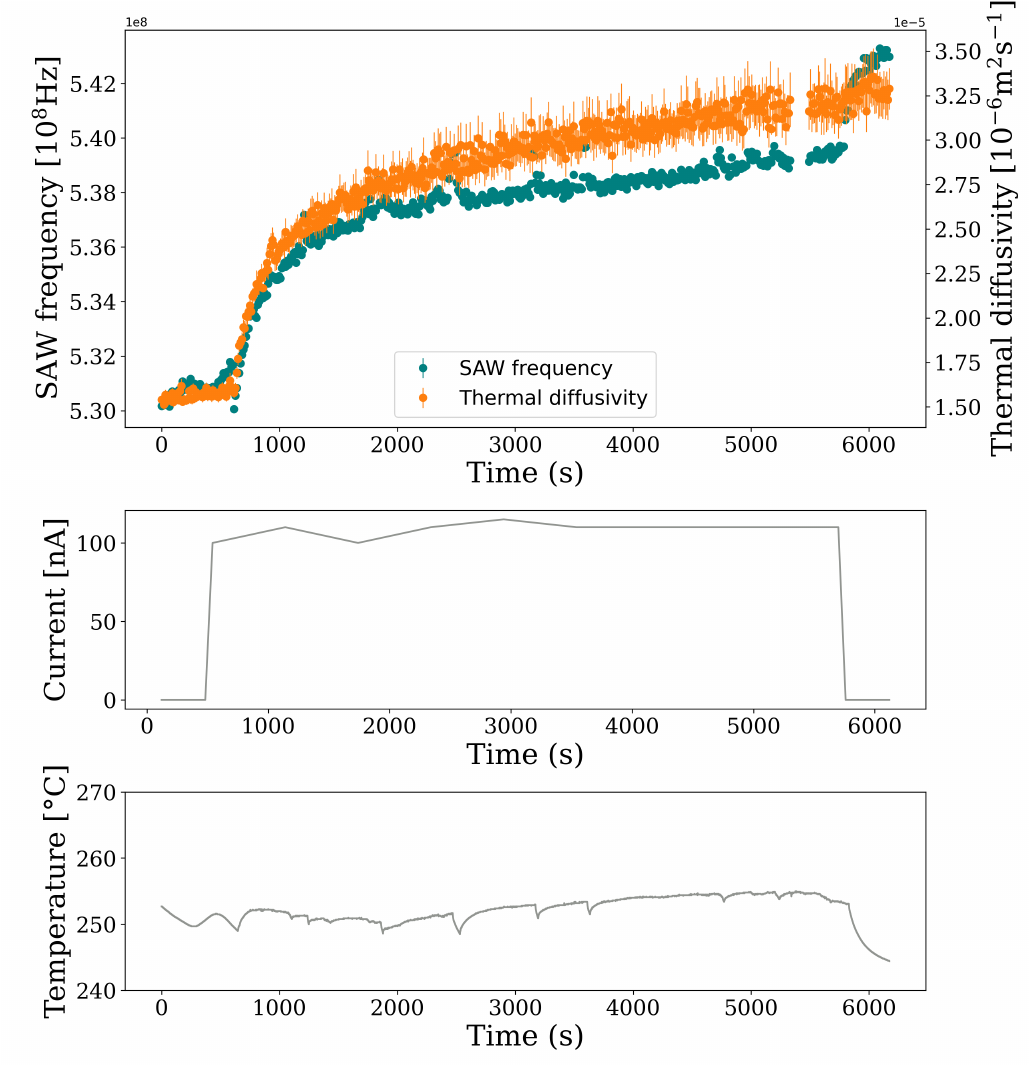}
    \caption{$I^3TGS$ results from the Cu–5 at.\% Cr–2.5 at.\% Ta sample irradiated at 6.5 MeV with Cu$^{3+}$ ions at $250^\circ C$. Blue dots correspond to SAW frequency values and orange dots to thermal diffusivity.}
    \label{fig:TGS_irr}
\end{figure}

Thermo-elastic properties were monitored during self-ion irradiation in 3 different samples of the Cu-Cr-Ta system, using the \textbf{I}n situ \textbf{I}on \textbf{I}rradiation experiments in combination with \textbf{TGS} measurements (\textbf{$I^3TGS$}). These type of measurements not only give information related to the evolution of the thermo-elastic properties of the material, but also allow to infer the relative radiation damage in the sample and compare it to other alloys manufactured under the same conditions \cite{botica-artalejo_real-time_2026} by monitoring the fluctuation of the SAW frequency. During these experiments samples have been irradiated continuously up to 25 DPA using 6.5 MeV Cu$^{3+}$ at $250^\circ C$ to promote vacancy clustering \cite{zinkle_void_1989} in order to facilitate void visualization, at the same time thick film integrity was guaranteed. When irradiation temperatures were close to $300^\circ C$, film delamination was observed during cool down, hence irradiation temperature was reduced to $250^\circ C$.
Figure \ref{fig:TGS_irr} shows the evolution of SAW frequency with time for a flux rate of $5.6\times 10^{16} ions/m^2\cdot s$ when using a beam current of $120nA$. Additionally, it can be seen the evolution of temperature and current during the whole irradiation time. 
It is noticeable a initial drop in the SAW frequency upon beam start, around 0.1-0.2 DPA. At this dose, some authors have identified the point defect saturation level for copper alloys after monitoring properties like electrical resistivity, thermal resistivity or hardening \cite{fabritsiev_effect_1996, singh_defect_1993, granberg_mechanism_2016, trachanas_study_2025}. A similar profile, with the same trend was obtained for the other samples in the present work. It's noticeable that the sample that shows the smallest drop of the SAW frequency, around 0.07 DPA, is the Cu–5 at.\% Cr–2.5 at.\% Ta. The values of the SAW frequency drop for each alloy can be seen on table \ref{tab:summary}.

\subsection{Radiation damage assessment through transmission electron microscopy}
\begin{figure}
    \centering
    \includegraphics[width=0.9\textwidth]{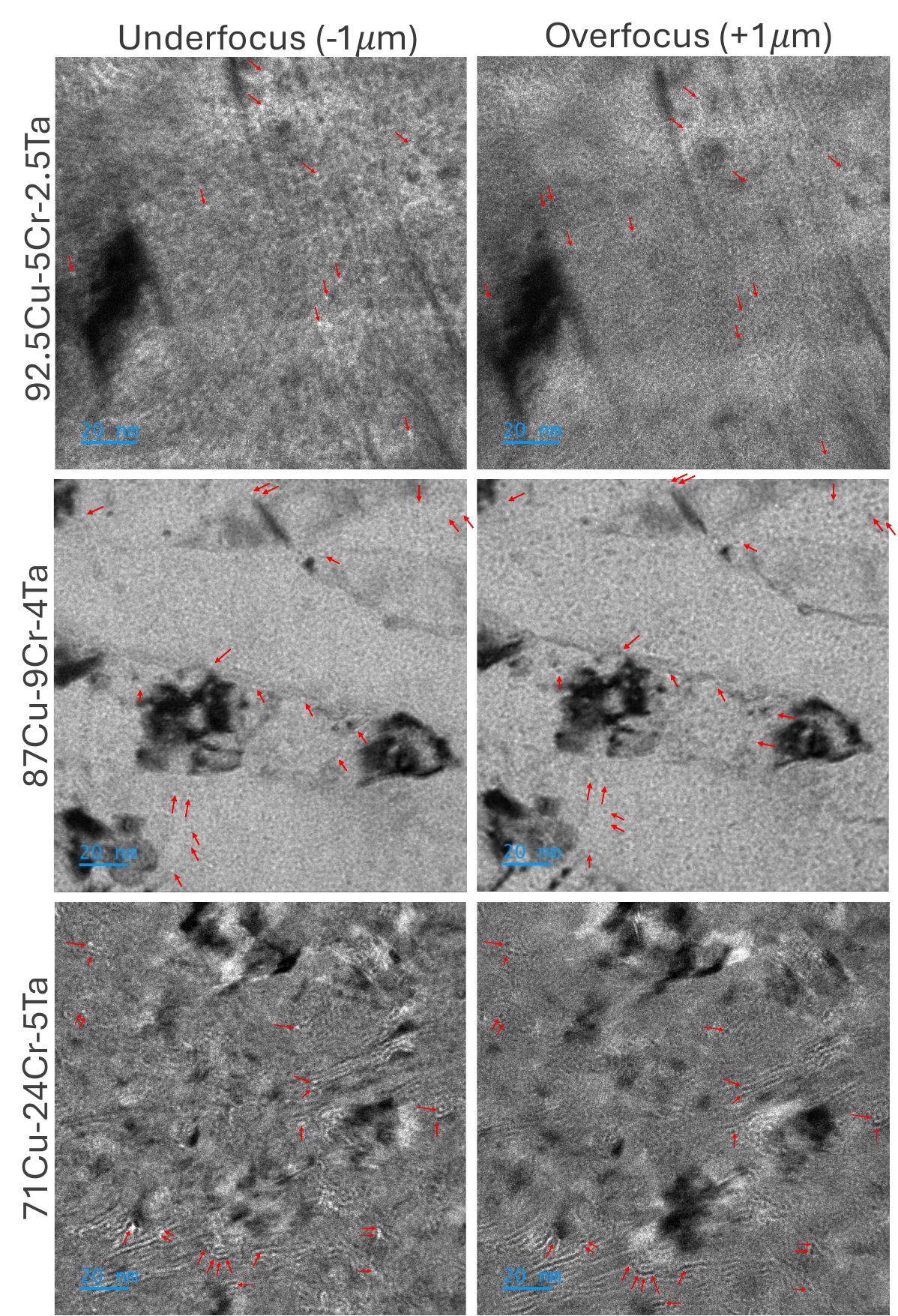}
    \caption{\small TEM images from the different Cu-Cr-Ta compositions. On the left side pictures were taken with an underfocus value of $-1\mu m$, and images on the right column were taken with an overfocus value of $+1\mu m$. Red arrows point out the voids.}
    \label{fig:TEM}
\end{figure}

\begin{figure}
    \centering
    \includegraphics[width=0.9\textwidth]{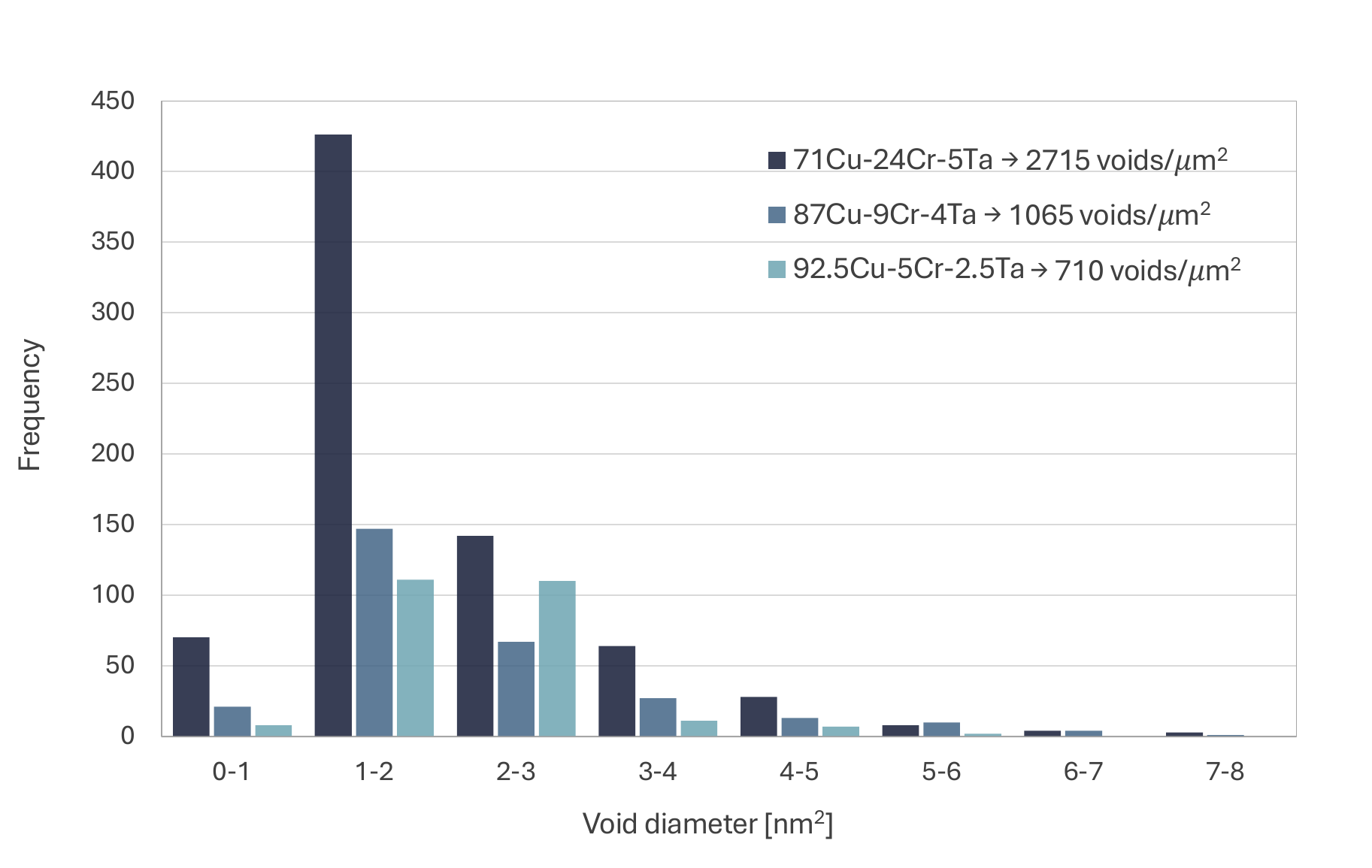}
    \caption{\small Void number and size distributions for each sample. Void density (voids per $\mu m ^2$) for each composition is shown in the upper-right corner.}
    \label{fig:void_dist}
\end{figure}

\begin{figure}
    \centering
    \includegraphics[width=0.9\textwidth]{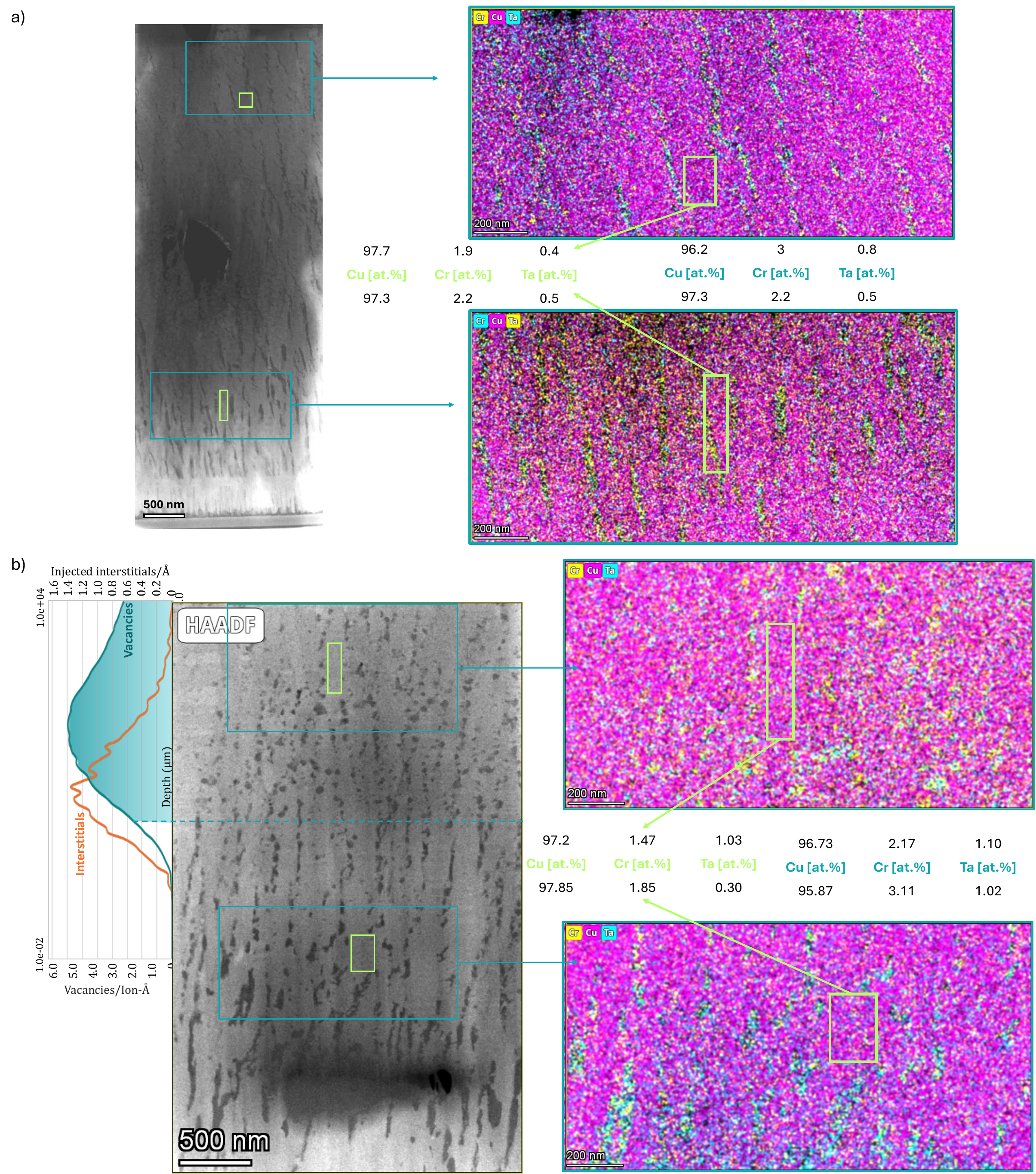}
    \caption{\small EDS analysis of sample 92.5Cu-5Cr-2.5Ta. a) shows a lamella od the material prior to irradiation and b) shows a lamella of the irradiated material next to the expected damage profile calculated with SRIM, which match the area of the material where no precipitates are found.}
    \label{fig:eds}
\end{figure}

To quantify radiation damage, the void density was determined from TEM images. All samples were examined in conventional TEM (CTEM) mode using the overfocus–underfocus technique \cite{williams_transmission_2008}, with defocus values of $\pm 1 \mu m$ relative to the focused condition (see Figure \ref{fig:TEM}). Figure \ref{fig:void_dist} summarizes the void size and number distributions for each sample, together with the corresponding void densities. The highest void density is observed in the Cu–24 at.\% Cr–5 at.\% Ta alloy, followed by Cu–9 at.\% Cr–4 at.\% Ta, while Cu–5 at.\% Cr–2.5 at.\% Ta exhibits the lowest void density.
A consistent trend is also observed in the SAW frequency drop, which increases with increasing void density. A direct comparison of the void density and SAW frequency drop trends is provided in Table \ref{tab:summary}.
To assess the effects of irradiation on the remaining microstructure, EDS was performed on both irradiated and unirradiated samples. Figure \ref{fig:eds} compares unirradiated (Fig. \ref{fig:eds}a) and irradiated (Fig. \ref{fig:eds}b) lamella extracted parallel to the irradiation direction from the Cu–5 at.\% Cr–2.5 at.\% Ta alloy. The unirradiated microstructure exhibits a homogeneous distribution of thin elongated $Cr_2Ta$ precipitates. In contrast, the irradiated lamella shows a marked reduction of these precipitates within the damage region, with precipitates predominantly observed beyond the penetration depth of the irradiation, indicating dissolution of the precipitates. Similar behavior is observed for the other alloy compositions.

\subsection{Statistics analysis of the vacancy migration energy barriers}
\begin{figure}
    \centering
    \includegraphics[width=0.9\textwidth]{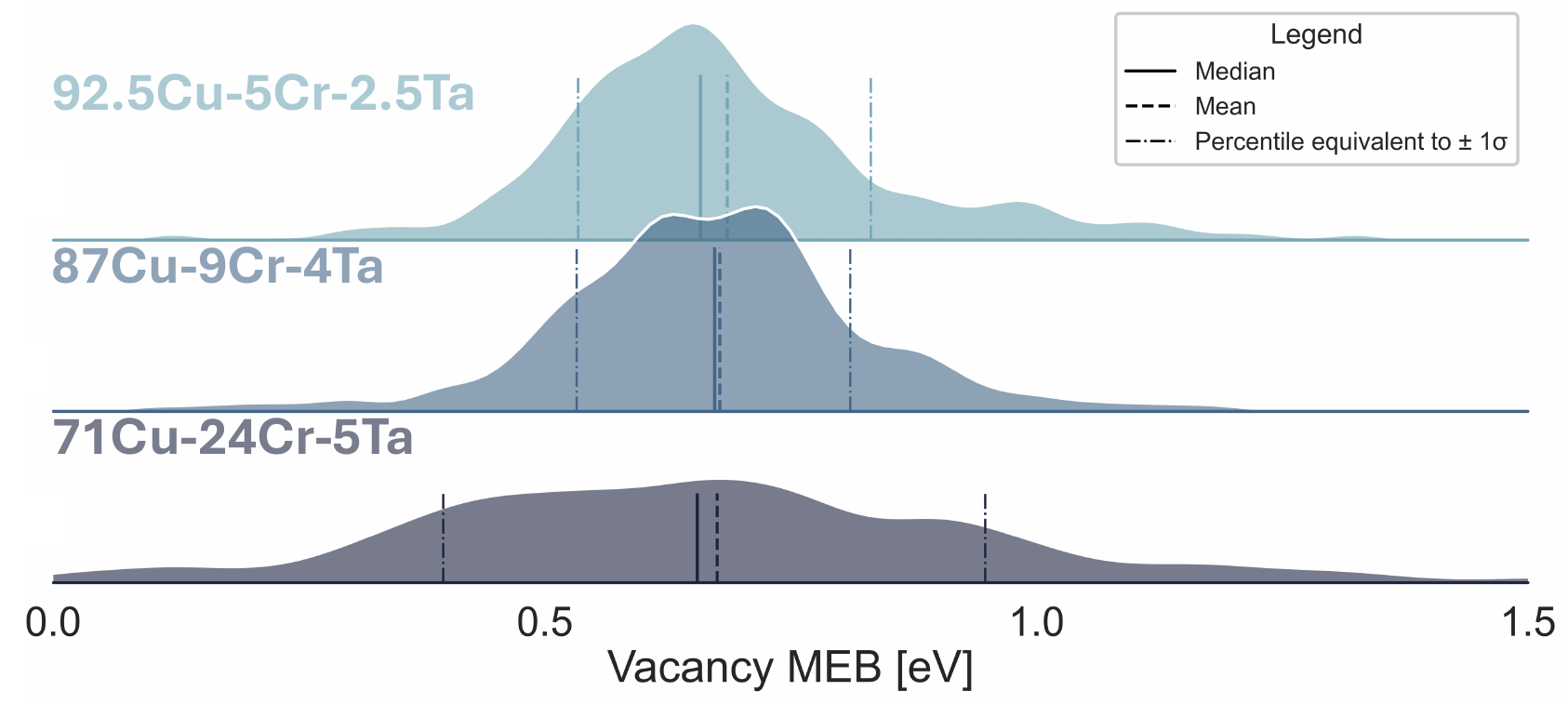}
    \caption{\small Distribution of the vacancy MEB for the different Cu-Cr-Ta compositions. The solid line denotes the median of the distribution, the dashed line indicates the mean, and the dash-dotted lines corresponds to the equivalent $\pm 1 \sigma$ interval.}
    \label{fig:MEB_dist}
\end{figure}

For each of the three Cu–Cr–Ta compositions that were experimentally tested and characterized, NEB calculations were performed to obtain the vacancy MEB with the aim of elucidating the influence of the potential energy landscape (PEL) on void formation propensity in each alloy. To obtain statistically meaningful distributions of vacancy MEBs, no explicit separation by atomic species was applied. Instead, an effective system-wide diffusivity was evaluated by considering the atomic species occupying the twelve nearest-neighbor sites surrounding a central vacancy in randomly generated simulation cells. Consequently, the resulting statistics inherently reflect the compositional abundance of each element and its likelihood of appearing in the local neighboring environment.
Figure \ref{fig:MEB_dist} shows violin plots for the vacancy MEB for each of the copper alloys. The median, mean, and percentile range equivalent to $\pm 1\sigma$ are indicated by solid, dashed, and dash-dotted lines, respectively.

\begin{table}[ht]
\centering
\caption{Summary of the different properties evaluated on the three systems of Cu alloys.}
\label{tab:summary}
\small
\begin{tabularx}{\textwidth}{>{\bfseries}c Y Y Y Y}
\toprule
\bfseries Technique & \bfseries Property & \bfseries Cu-5at.\%Cr-2.5at.\%Ta & \bfseries Cu-9at.\%Cr-4at.\%Ta & \bfseries Cu-24at.\%Cr-5at.\%Ta \\
\midrule
\bfseries TGS   & $f_{SAW}$ drop [\%] (@DPA) & -0.30 (0.07) & -0.54 (0.07) & -1.27 (0.19) \\
\hdashline
\bfseries TEM   & Void density $[\#/\mu m^2]$ & 710 & 1065 & 2715 \\
\hdashline
\multirow{4}{*}{\bfseries NEB} 
                  & ${\eta}_{MEB} [eV]$ & 0.658 & 0.673 & 0.655 \\
                  & ${\mu}_{MEB} [eV]$ & 0.685 & 0.678 & 0.675 \\
                  & ${\sigma}_{MEB} [eV]$ & 0.184 & 0.197 & 0.304 \\
                  & $LB{\sigma}_{MEB} [eV]$  & 0.534 & 0.532 & 0.397 \\
\bottomrule
\end{tabularx}
\end{table}

\section{Discussion}
\subsection{Alloy system selection}
CALPHAD predictions of thermal conductivity evolution with temperature on different Cu-Cr-X alloy systems indicate that tantalum is the most effective alloying element for RF antenna materials. In addition to preserving favorable thermal and electrical transport properties, Ta promotes precipitation strengthening through the formation of C15 Laves phase precipitates dispersed within the copper matrix. These combined characteristics are particularly relevant for applications requiring both high heat removal capability and mechanical stability under service conditions.
Beyond thermophysical performance, radiological considerations play a critical role in alloy selection for RF antennas for fusion reactors. The allowable Ta concentration range defined by the Fetter limits (6–8 at.\%) \cite{fetter_long-term_1990} provides sufficient compositional flexibility to achieve meaningful precipitate strengthening while maintaining a low activation response under fusion spectrum neutron irradiation. When considered alongside the CALPHAD results, these constraints highlight the Cu–Cr–Ta system as a well-balanced solution, offering a favorable compromise between thermal performance, microstructural stability, and radiological viability. Consequently, Cu–Cr–Ta emerges as a strong candidate system for fusion-relevant components such as RF antennas, diagnostics, and normally-conducting magnet coils which may experience significant neutron irradiation.

\subsection{Primary radiation damage resistance comparison}
As discussed previously, copper-based alloys exhibit saturation of point defects at damage levels of approximately 0.1–0.2 DPA. This behavior is also observed in the three alloys investigated in this study (see Table \ref{tab:summary}). The evolution of the SAW frequency therefore provides a direct means of tracking changes in material compliance as point defects accumulate during irradiation, considering that vacancies are the main contributor to the thermo-elastic properties of the material, due to their slower diffusivity \cite{sivak_diffusion_2022}. Once point defect saturation is reached, additional hardening mechanisms, such as interactions with pre-existing dislocations, between others, become dominant \cite{dennett_dynamic_2021}, facilitating the identification of this inflection point.

The magnitude of the SAW frequency decrease between the unirradiated state and the saturation point (prior to the subsequent increase in SAW frequency) reflects the extent to which the material softens due to vacancy accumulation. A larger frequency drop indicates a greater increase in compliance, implying a higher retained vacancy population in the microstructure once the equilibrium concentration of point defects has been established. Assuming similar defect evolution mechanisms across the alloys from the compositional space explored, those that generate a larger population of point defects during the primary damage stage are expected to exhibit an increased susceptibility to void formation as irradiation progresses, implying a larger influence on material's properties.

Conversely, a smaller SAW frequency drop up to the point defect saturation regime indicates reduced vacancy accumulation and, consequently, greater resistance to primary radiation damage. Among the compositions examined here, the Cu–5 at.\% Cr–2.5 at.\% Ta alloy exhibits the smallest SAW frequency decrease, suggesting superior resistance to primary radiation damage.

This interpretation is further supported by post-irradiation TEM analyses conducted on the $I^3TGS$ samples exposed up to 25 DPA, which reveal trends in void density consistent with those inferred from the SAW frequency decay measurements. Moreover, ratios of void densities between pairs of alloys (e.g., Void density$_{92.5Cu-5Cr-2.5Ta}$ / Void density$_{71Cu-24Cr-5Ta}$) agree with the corresponding SAW frequency decay ratios to within ±0.05. This agreement demonstrates that $I^3TGS$ enables reliable \textit{relative} comparisons of primary radiation damage resistance among alloys fabricated under identical conditions. As such, $I^3TGS$ provides an efficient tool for early-stage compositional down-selection, significantly reducing the need for accelerator time and extensive post-irradiation characterization.

\subsection{Radiation-induced void formation explained by vacancy MEB distributions}
Several studies have shown that the roughness of the PEL in chemically complex alloys can suppress defect mobility and enhance primary radiation damage resistance. Jin et al. \cite{jin_thermodynamic_2018} established a link between system mixing energy, heterogeneity in defect MEBs, and radiation tolerance in NiFe solid solutions, demonstrating via molecular dynamics that compositions with larger MEB variance exhibited reduced primary damage. %Similar correlations were reported by Segantin and Short for W–Ta and V–Cr alloys, where increased MEB variance consistently corresponded to lower predicted radiation damage \cite{?}.%
On the other hand, Xu et. al \cite{xu_mechanism_2023} claim that in multi-principal element alloys (MPEAs), the mean value of the MEB of each elemental species has a stronger influence on diffusion behavior than the standard deviation of the barrier distribution. This is because the likelihood of an atom making a jump is determined mainly by its average barrier height, while variations around that mean have a comparatively smaller effect. Interestingly, Bin et al. \cite{bin_short-range_2022} showed that short-range order (SRO) in MPEAs distorts the PEL, producing asymmetric forward and backward migration barriers that act as trapping sites and suppress long-range diffusion.
In parallel, high mixing energy has been associated with a reduced propensity for large defect cluster formation following successive radiation cascades \cite{jin_thermodynamic_2018}, while slow energy dissipation has been proposed as an additional governing mechanism controlling early-stage defect dynamics \cite{zhang_influence_2015}. Experimental and simulation work by Granberg et al. \cite{granberg_mechanism_2016} further demonstrated that equiatomic alloys exhibit enhanced resistance to irradiation-induced amorphization, attributed to lattice distortion that limits dislocation motion and constrains defect cluster growth. Together, these studies indicate that chemical complexity could enhance radiation tolerance through multiple, interconnected mechanisms, all of which ultimately influence defect mobility, recombination, and clustering.

To interpret the experimental observations in the present work, vacancy migration energy barrier calculations were performed for each irradiated composition, and statistical descriptors of the MEB distributions (Figure \ref{fig:MEB_dist}) are shown in Table \ref{tab:summary}. No direct correlation was found between either the mean or the standard deviation of the vacancy MEB distributions and the void densities measured by TEM. Additionally, as we considered the nearest 12 neighboring positions of the FCC and calculate the vacancy MEB independently of the element on that position, to account for the compositional abundance of each element and its likelihood of appearing nearby to that random vacancy, non-gaussian distributions of the MEB were obtained. Then, as it is an asymmetrical distribution, the median is a more representative value than the mean, in the same way the percentiles equivalent to the $\pm 1 \sigma$ of the population have been calculated. Therefore, the analysis of the lower tail of the vacancy MEB distributions reveals a physically meaningful trend. Alloys exhibiting smaller vacancy MEB values at the $-1\sigma$ level possess a higher fraction of vacancies capable of escaping the cascade region, as the energy that they have to overcome is lower, thereby reducing vacancy–interstitial recombination during continuous irradiation and increasing the probability of void formation in the long term irradiation. Conversely, compositions with larger $-1\sigma$ MEB values promote more localized vacancy populations, as the diffusivity of those vacancies will take longer, enhancing recombination with cascade-generated interstitials and suppressing void nucleation. These results suggest that radiation damage resistance is governed by the fraction of highly mobile vacancies that control recombination efficiency and defect survival.

\subsection{Precipitate recoil dissolution}
TEM imaging over the full lamella length, before and after self-ion irradiation, reveals clear dissolution of precipitates within the irradiated region. This behavior is noteworthy, as previous studies on Cu–Cr–Nb and Cu–Cr–Nb–Zr alloys subjected to heavy-ion and neutron irradiation did not report precipitate dissolution \cite{seltzman_nuclear_2020, perrin_microstructure_2024}. A key distinction between the present samples and those investigated in earlier work lies in the characteristic precipitate size relative to the matrix grain size. While precipitates in the previously reported alloys are typically much smaller than the matrix, the precipitates in this study have a comparable size respect the matrix grains.

The observed behavior can be rationalized in terms of recoil dissolution, in which the relative size of the precipitates with respect to the surrounding matrix grain size governs their stability under irradiation \cite{was_fundamentals_2017}. When precipitate dimensions approach those of the matrix grains, as in the present microstructure, ballistic mixing promotes radiation-induced dissolution. In contrast, smaller precipitates embedded in coarse-grained matrices are more resistant to recoil-driven dissolution and may instead remain stable or even grow under irradiation. Consequently, although precipitate dissolution is observed in the nanostructured PVD samples studied here, such behavior is not expected for bulk Cu–Cr–Ta alloys subjected to similar irradiation conditions.

\section{Conclusion}
Considering the current limitations in material availability for RF launchers in fusion reactors, the Cu–Cr–Ta alloy system emerges as a promising candidate due to its ability to form C15 Laves phase precipitates, which provide mechanical stability at elevated temperatures while preserving favorable thermal and electrical conductivity and maintaining acceptable levels of WDR. CALPHAD-based screening further supports the selection of Ta as an optimal alloying element, offering improved thermal conductivity performance relative to established benchmarks such as GRCop-84 while satisfying radiological constraints relevant to fusion environments.

In this work, a novel approach for evaluating primary radiation damage resistance was demonstrated using in situ ion irradiation combined with transient grating spectroscopy ($I^3TGS$). By monitoring the evolution of the surface acoustic wave (SAW) frequency from the unirradiated state to the point defect saturation regime, relative differences in vacancy accumulation among alloys could be quantified in real time. Post-irradiation transmission electron microscopy of samples exposed up to 25 DPA validated this methodology, revealing a strong correlation between the magnitude of the SAW frequency drop and the resulting void density. This agreement confirms that $I^3TGS$ enables reliable, rapid comparison of radiation damage resistance between alloys fabricated under identical conditions, significantly reducing the reliance on long irradiation campaigns and extensive post-irradiation characterization during early-stage compositional down-selection.
Among the investigated compositions within the Cu–Cr–Ta system, the Cu-rich alloy (Cu–5 at.\% Cr–2.5 at.\% Ta) exhibited the smallest SAW frequency decrease during irradiation, as well as the lowest void density, indicating superior resistance to radiation. Complementary vacancy migration energy barrier analyses suggest that this enhanced performance is associated with a reduced population of highly mobile vacancies, promoting vacancy–interstitial recombination and decreasing void formation.

%% The Appendices part is started with the command \appendix;
%% appendix sections are then done as normal sections
%%\appendix
%%\section{Example Appendix Section}
%%\label{app1}

\section*{Data availability statement}
The data and code that support the findings of this study are openly available in the following repository: https://github.com/shortlab/2026-ebotica-CuCrTa-RDRA.

\section*{Declaration of generative AI and AI-assisted technologies in the manuscript preparation process}
During the initial preparation of this work, ChatGPT-5 model were used by the main author for English correction and expression. After using this tool, all authors reviewed and edited the content as needed, and take full responsibility for the content of the publication.  

\section*{CRedit authorship contribution statement}
\textbf{Elena Botica-Artalejo} Conceptualization, Methodology, Formal Analysis, Investigation, Writing – Original Draft, Writing – Review \& Editing, Visualization.  \textbf{Gregory M. Wallace} Conceptualization, Methodology, Supervision, Funding acquisition, Writing - Review \& Editing. \textbf{Michael P. Short} Conceptualization, Methodology, Supervision, Writing - Review \& Editing.

\section*{Declaration of Competing Interest} 
The authors declare that they have no known competing financial interests or personal relationships that could have appeared to influence the work reported in this paper.

\section*{Funding sources}
This work was supported by the U.S. Department of Energy, Office of Science, Office of  Fusion Energy Sciences under Award Number DE-SC0024307.

\section*{Acknowledgment}
This work was performed in part in the MIT.nano Characterization Facilities and in the Center of Nanoscale Systems at Harvard University. Thanks to Aubrey Penn at MIT.nano for her help with the Themis Z G3 Cs-Corrected S/TEM.

%\bibliographystyle{elsarticle-num} 
%\bibliography{references}
\printbibliography
\end{document}